\documentclass[slac_one]{revtex4}
\usepackage{graphicx}
\usepackage{fancyhdr}
\newcommand{\met}{\mbox{\ensuremath{\slash\kern-.7emE_{T}}}}
\newcommand{\invfb}{fb$^{-1}$}

\begin{document}

\title{Searches beyond the standard model at high-energy colliders} 

%

\author{J.-F. Grivaz}
\affiliation{Laboratoire de l'Acc\'el\'erateur Lin\'eaire, Orsay, France}

\begin{abstract}
Recent searches for physics beyond the standard model at high-energy colliders 
are reviewed, with emphasis on supersymmetry, additional space dimensions, 
extra gauge bosons, leptoquarks, and model-independent searches.
The results reported are based on data samples of up to 0.5 and 2.5~\invfb\
collected at HERA and at the Tevatron, respectively.
\end{abstract}

\maketitle

\thispagestyle{fancy}


\section{INTRODUCTION} 
This review covers results presented in three parallel sessions (SUSY, and 
Exotics I and II) during which 21 experimental talks were delivered, of which
ten reported results from the Tevatron, four from HERA, one from the B 
factories, and six gave prospects at the LHC. In addition, there were twelve
theoretical talks, for which the reader should consult Refs.~\cite{CW} and 
\cite{GB}. Furthermore, in view of the limited amount of time allocated to this
review, hard choices had to be made, for which I apologize. In particular,
only actual results, i.e., no prospects, will be reported here.

There are many ways in which the standard model (SM) can be extended. 
Supersymmetry (SUSY) 
is the only non-trivial extension of the Poincar\'e group, and it naturally
incorporates gravity in its local form. String theories suggest that there 
could be a number of compact space dimensions in addition to the three usual 
ones.
The SU(3)$\times$SU(2)$\times$U(1) gauge group can be seen as resulting from
the breaking of a larger, more unified, gauge group. Alternatives
to the Higgs mechanism can be considered to break the electroweak symmetry.
The quark-and-lepton structure in three generations is at the origin of many 
ideas such as compositeness.

Among those many possibilities, the choice was made in this review to address
supersymmetry, additional space dimensions, new gauge bosons, and leptoquarks.
Searches for new phenomena not motivated by any specific model will also be
reported. In the following, all limits will be given at 95\% C.L., and 
evidences or discoveries will be reported only if their significance exceeds
three or five sigmas, respectively. 

\section{SUPERSYMMETRY}
The general concepts of supersymmetry, as well as details of the minimal
supersymmetric extension of the standard model, the MSSM, can be found in
Ref.~\cite{CW}. Briefly, the MSSM is built from 
the minimal field content, with squarks,
sleptons, gluinos, charginos and neutralinos, and with two Higgs doublets.
In the absence of a specific SUSY breaking scheme, more than a hundred new
parameters are introduced, which are reduced to a manageable level by invoking
some unification of the soft SUSY-breaking terms at the GUT scale. In what is
usually called ``standard SUSY'', additional assumptions are made: in order to 
avoid fast proton decay, $R$-parity is assumed to be conserved, and the 
lightest supersymmetric partner, the LSP, is a neutralino. As a consequence,
SUSY particles are produced in pairs, and they decay to SM 
particles and to the LSP; since the LSP is stable and weakly interacting, the 
final state exhibits missing energy. Models in which $R$-parity is not 
conserved are not considered in this review.

The most widely studied model is minimal supergravity, mSUGRA. Only five 
parameters are needed to fully specify the model: a common scalar mass $m_0$,
a common gaugino mass $m_{1/2}$, and a common trilinear coupling $A_0$, all
defined at the GUT scale, the ratio $\tan\beta$ of the vacuum expectation 
values of the two Higgs doublets, and the sign of $\mu$, the supersymmetric
Higgs mass term. The absolute value of $\mu$ is determined from the 
renormalization group equations, by imposing electroweak symmetry breaking at
the proper scale. The typical mSUGRA spectrum contains a light Higgs boson, 
a neutralino LSP, sleptons substantially lighter than squarks, and the 
following relation holds approximately among the gaugino masses: 
$m_{\tilde{g}}/3\sim m_{\tilde\chi^\pm}\sim 2m_{\tilde\chi_1^0}$.

\subsection{Standard SUSY}
Some of the most constraining limits on SUSY particles still result from the
searches performed at LEP. Mass limits of about 100~GeV were obtained for 
sleptons 
and charginos~\cite{lepslepchar}.
In the MSSM with sfermion and gaugino mass unification, a mass lower limit 
of 47~GeV was set on the LSP at large $\tan\beta$~\cite{leplsp}, 
somewhat tighter at lower $\tan\beta$ and in mSUGRA~\cite{lepsugra}.


There are two main streams for standard-SUSY searches in $p\bar{p}$
collisions at the Tevatron: squarks and gluinos in jets+\met\ final states, and
electroweak gauginos in the trilepton topology.

\subsubsection{Squarks and gluinos}
Colored SUSY particles, squarks and gluinos, are expected to be the most 
copiously produced at the Tevatron via the strong interaction, up to phase 
space suppressions at large masses. If squarks are much lighter than gluinos,
the most relevant production channel is $p\bar{p}\to\tilde{q}\tilde{\bar{q}}$.
In that configuration, the $\tilde{q}\to q\tilde\chi_1^0$ decay mode is
expected to be dominant. If gluinos are much lighter than squarks, the 
$p\bar{p}\to\tilde{g}\tilde{g}$ production channel and the 
$\tilde{g}\to q\bar{q}\tilde\chi_1^0$ decay mode are expected to be the most
relevant. If the squark and gluino masses are similar, both production
processes contribute, as well as squark-gluino associated production,
$p\bar{p}\to\tilde{q}\tilde{g}$. The above processes lead to topologies of
at least two, four, or three jets with \met, respectively, and the CDF and D\O\
collaborations have both optimized three analyses accordingly, the results of 
which are combined in the end. The simple picture outlined above is in practice
complicated by the possibility of cascade decays such as 
$\tilde{q}\to q'\tilde\chi^\pm\to q'\ell\nu\tilde\chi_1^0$, so that a specific
model is needed to interpret the search results. 

\begin{figure}
\begin{tabular}{cc}
\includegraphics[width=8.5cm]{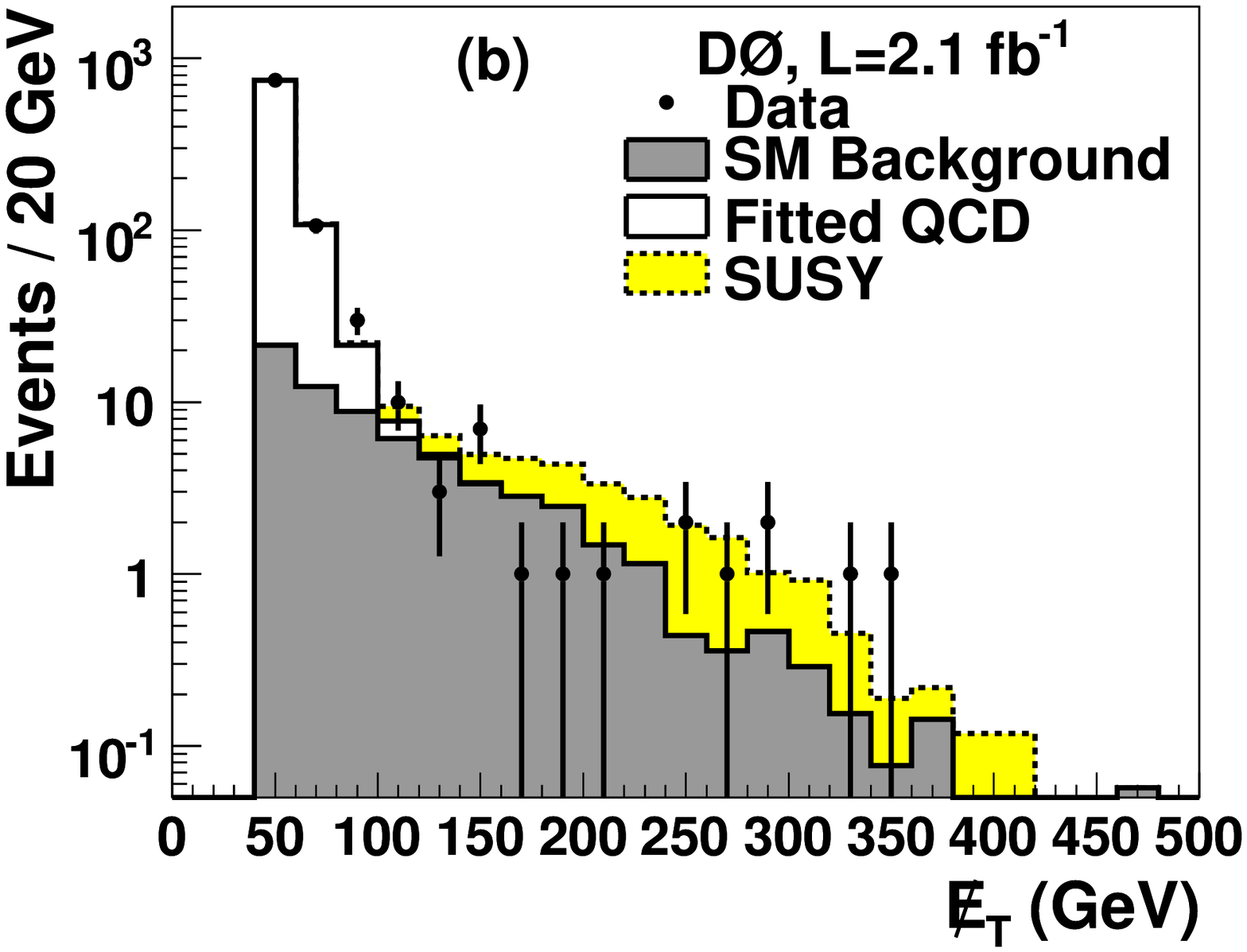} &
\includegraphics[width=8.5cm]{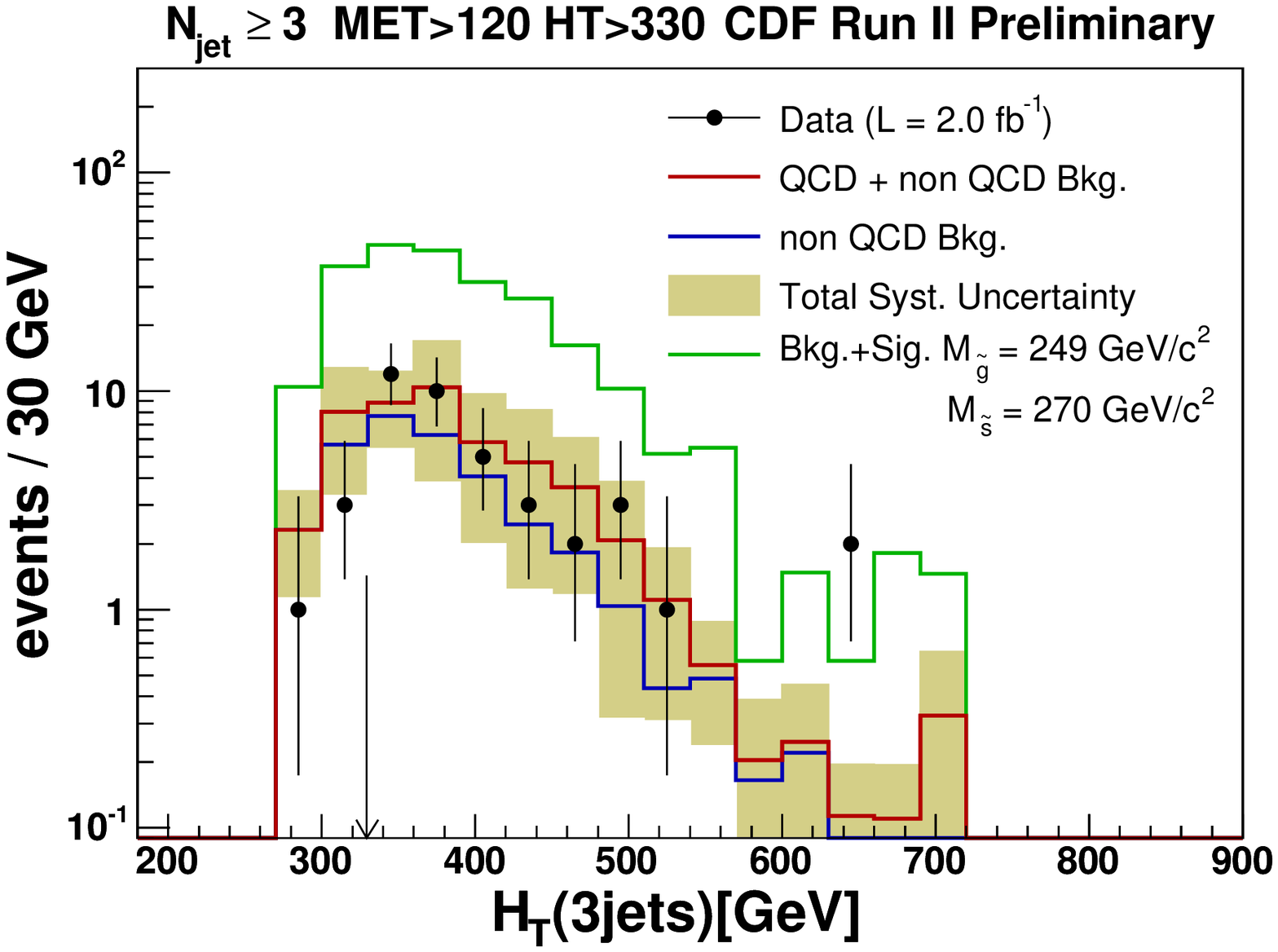} 
\end{tabular}
\caption{\label{methtsqgl}
Left: Final \met\ distribution in the D\O\ search for squarks and 
gluinos~\cite{DZsqgl} in the three-jet topology, after $H_T>375$~GeV; 
the final cut is $\met>175$~GeV. 
Right: Final $H_T$ distribution in the CDF search for squarks and 
gluinos~\cite{CDFsqgl} in the three-jet topology, after $\met>120$~GeV; 
the final cut is $H_T>330$~GeV.}
\end{figure}

In the searches in jets+\met\ topologies, the main backgrounds are of two 
types: instrumental from standard QCD multijet production, where the \met\
arises from jet energy mismeasurements, and $(W/Z)$+jets production, with
real \met\ due to the neutrinos from $W\to\ell\nu$ or $Z\to\nu\bar\nu$. The
instrumental background is reduced by the requirement that the \met\ should 
not be aligned with the direction of a jet, while the $W$+jets background 
(including from top quark decays) is 
reduced by vetoing events with isolated leptons. The background from 
$(Z\to\nu\bar\nu)$+jets is irreducible. The final discrimination is achieved 
by requiring large \met\ and $H_T$, where $H_T$ is the scalar sum of the jet 
transverse momenta. Final distributions of these variables in the CDF and D\O\ 
three-jet analyses are shown in Fig.~\ref{methtsqgl}.  

\begin{figure}
\begin{tabular}{cc}
\includegraphics[width=8.5cm]{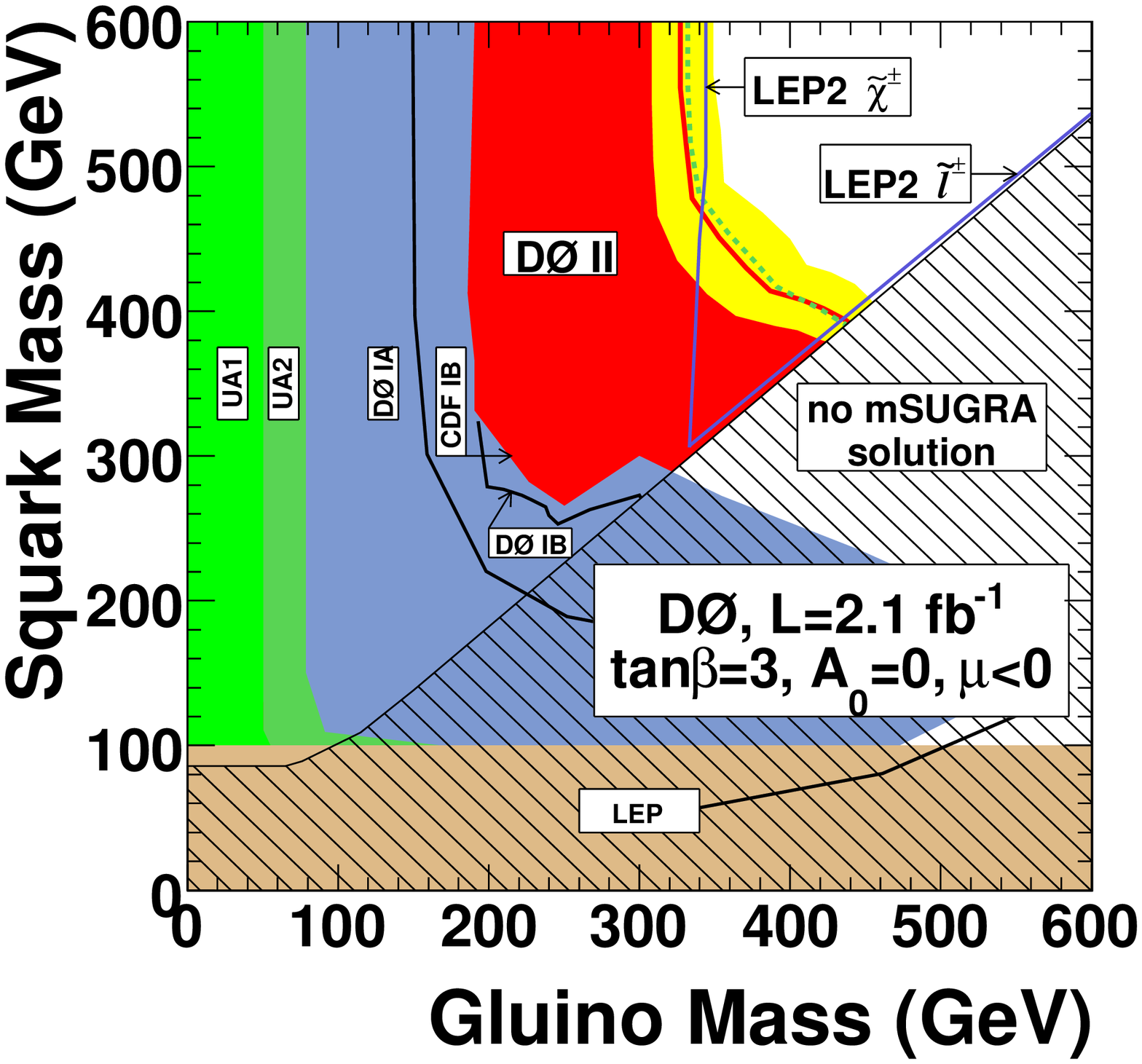} &
\includegraphics[width=8.5cm]{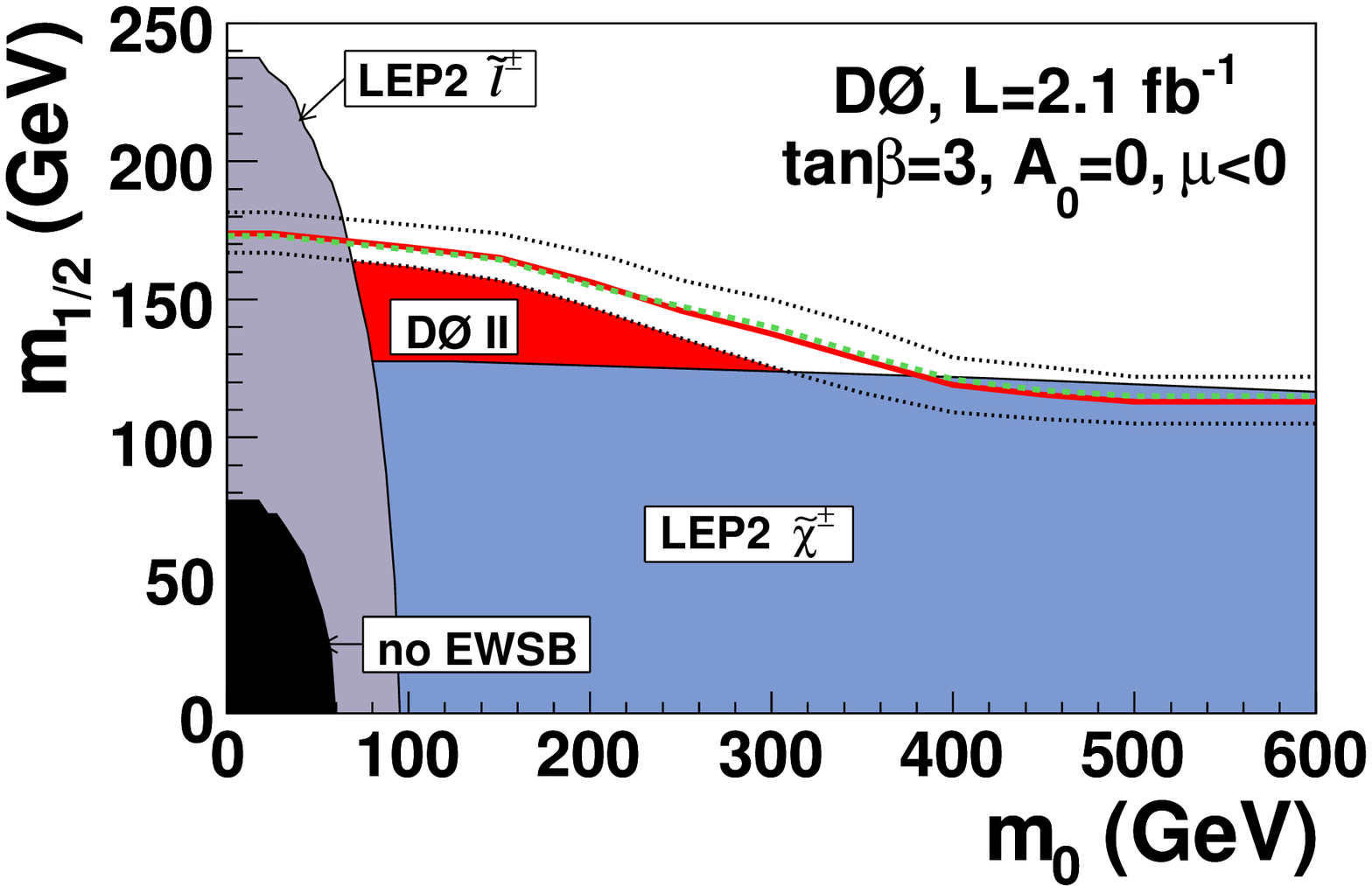} 
\end{tabular}
\caption{\label{sqgl}
Left: Region in the ($m_{\tilde{g}},m_{\tilde{q}}$) plane excluded by 
D\O~\cite{DZsqgl}
and by earlier experiments. The red curve corresponds to the nominal scale and 
PDF choices. The yellow band represents the uncertainty associated with these
choices. The blue curves represent the indirect limits inferred from the LEP
chargino and slepton searches.
Right: Region in the ($m_0,m_{1/2}$) plane excluded by the D\O\ search for 
squarks and gluinos~\cite{DZsqgl} and by the LEP experiments.}
\end{figure}

In the D\O\ analysis~\cite{DZsqgl} based on 2.1~\invfb, 
11, 9, and 20 events were selected 
in the two, three, and four-jet topologies, respectively, to be compared with 
background expectations of $11.1\pm 2.9$, $10.7\pm 2.8$, and $17.7\pm 4.5$ 
events, respectively. These results were turned into an exclusion domain in the
$(m_{\tilde{g}},m_{\tilde{q}})$ plane, within the mSUGRA framework with
$\tan\beta=3$, $A_0=0$, and $\mu<0$, as shown in Fig.~\ref{sqgl}-left. 
The exact
boundary of the exclusion domain depends on the choice of renormalization and
factorization scale, and on the parton distribution functions chosen, as shown 
by the yellow band in Fig.~\ref{sqgl}-left. 
For the nominal choice, indicated by the
red curve, squark masses below 392~GeV and gluino masses below 327~GeV are
excluded. The lower limit for equal squark and gluino masses is 408~GeV. An
alternative presentation can be given in terms of the mSUGRA parameters, as
shown in Fig.~\ref{sqgl}-right, 
where it can be seen that the domain excluded by 
gaugino and slepton searches at LEP is extended. Similar results have been
obtained by the CDF collaboration~\cite{CDFsqgl}.

\subsubsection{Electroweak gauginos}
Although chargino and neutralino production in $p\bar{p}$ collisions is 
mediated by electroweak interactions, 
while squarks and gluinos are produced via 
the strong interaction, the search for electroweak gauginos at the Tevatron is
nevertheless quite relevant for two reasons. In a framework with gaugino mass 
unification, a similar fraction of the parameter space is probed for 
electroweak-gaugino
masses a factor of three smaller than the gluino mass, thus reducing the impact
of the phase space suppression. Second, if sleptons are sufficiently light, 
leptonic decays of the electroweak gauginos such as 
$\tilde\chi^\pm\to\tilde\ell^\ast\nu\to\ell\nu\tilde\chi_1^0$ and 
$\tilde\chi_2^0\to\tilde\ell^\ast\ell\to\ell^+\ell^-\tilde\chi_1^0$ 
are enhanced. As a result, 
associated $\tilde\chi^\pm\tilde\chi_2^0$ production leads to the very clean 
signature of three leptons and \met\ in an otherwise quiet environment. The
final state leptons tend however to carry little energy, because the LSP mass
is roughly half the chargino or $\tilde\chi_2^0$ mass.

The search strategy is therefore to require two isolated leptons, with minimum
transverse momenta as low as possible, and some \met\ due to the escaping 
neutrino and neutralinos. The third lepton can either be explicitly identified,
or simply be an isolated charged particle track, thus providing some 
sensitivity to hadronic $\tau$ decays. Since this third lepton may be too soft
to be part of the analysis, an alternative strategy is also used, based on only
two identified leptons, but with the requirement that they should  have the 
same electric charge, a configuration for which the SM backgrounds 
are greatly reduced.

\begin{figure}
\begin{tabular}{cc}
\includegraphics[width=8.5cm]{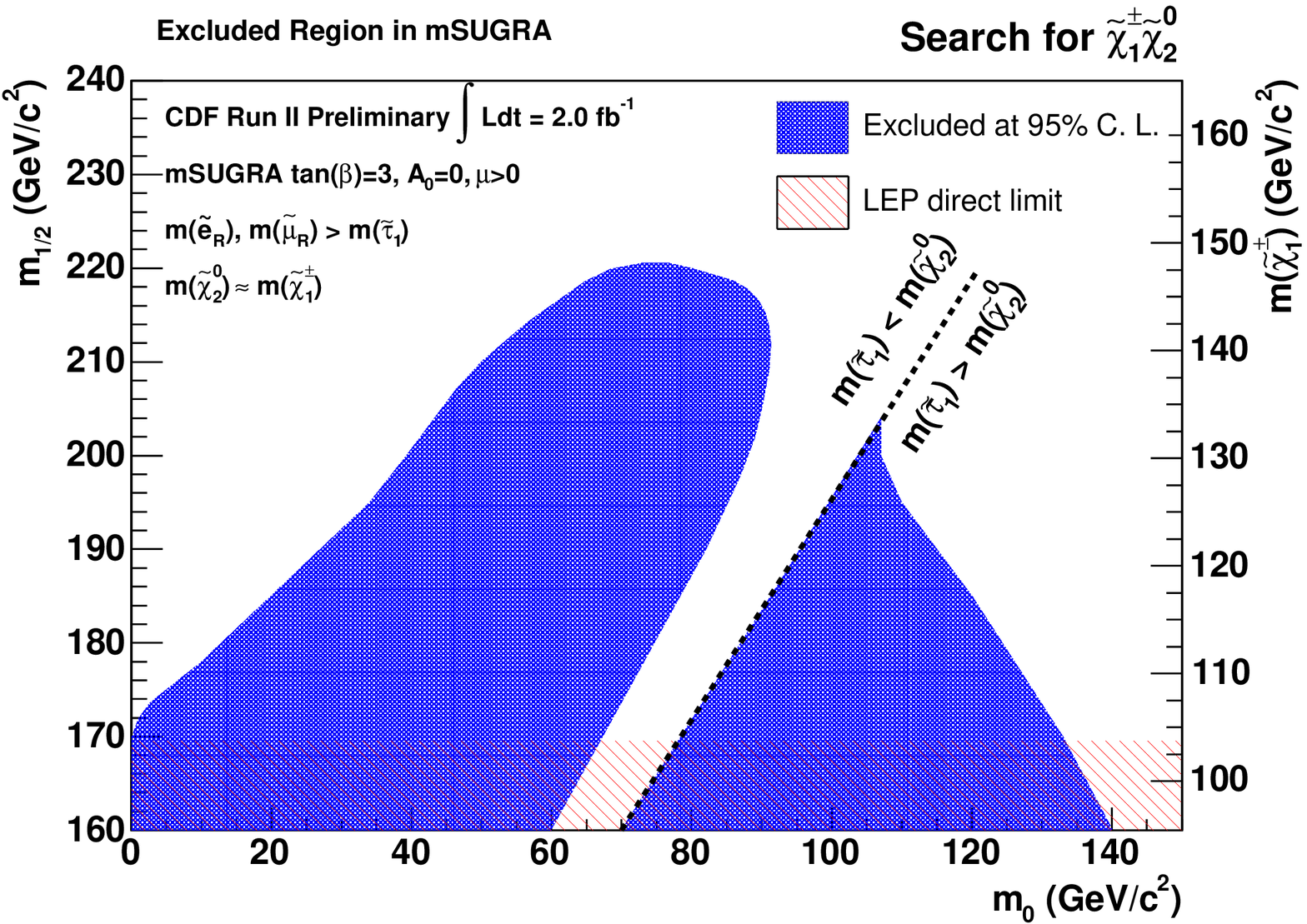} &
\includegraphics[width=8.5cm]{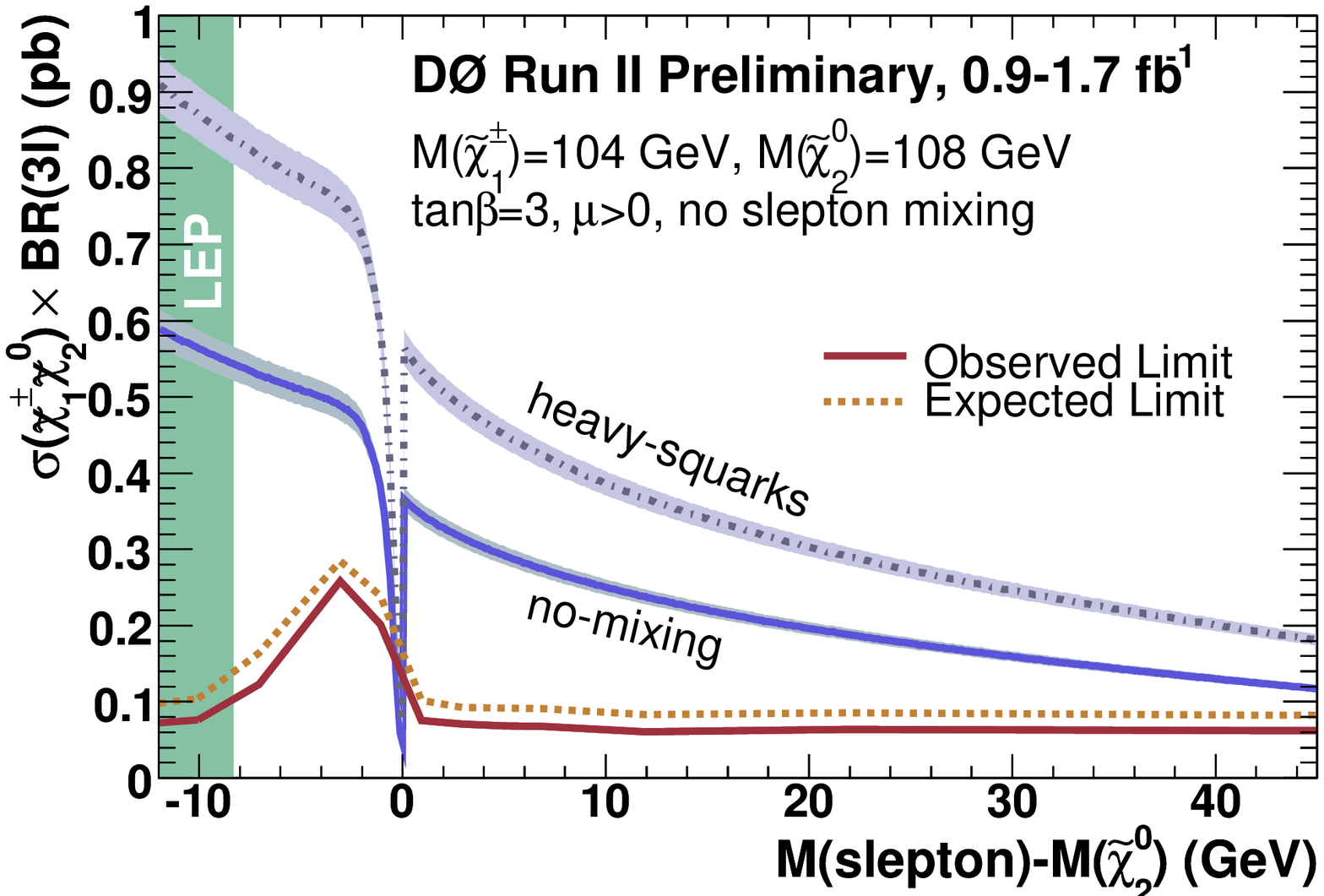}
\end{tabular}
\caption{\label{tril}
Left: Regions in the ($m_0,m_{1/2}$) plane excluded 
by the CDF search for trileptons~\cite{CDFtril}.
Right: Upper limit obtained by D\O\ on the product of the cross section of 
associated $\tilde\chi^\pm\tilde\chi_2^0$ production times the branching 
fraction into three leptons, as a function of the slepton -- $\tilde\chi_2^0$ 
mass difference, compared with the MSSM expectation without slepton mixing
(``no-mixing'' curve)~\cite{DZtril}.}
\end{figure}

In the CDF search~\cite{CDFtril} based on 2.0~\invfb, one trilepton event 
survived all stages of the selection, with a background expectation of 
$0.88\pm 0.14$ events. In the less pure selection where the third lepton is 
identified simply as an isolated track, six events were selected, with
$5.5\pm 1.1$ expected from backgrounds. These results were translated into
an exclusion domain in the mSUGRA parameter space, as shown in 
Fig.~\ref{tril}-left
for $\tan\beta=3$, $A_0=0$, and $\mu>0$. Two lobes can be observed. The left 
one corresponds to two-body decays such as $\tilde\chi^\pm\to\tilde\ell\nu$, 
while the right one corresponds to three-body decays such as 
$\tilde\chi_2^0\to\ell^+\ell^-\tilde\chi_1^0$. For $m_0=60$~GeV, a chargino 
mass lower limit of 145~GeV is obtained; the limit is 127~GeV for 
$m_0=100$~GeV. These values exceed the chargino mass lower limit set at LEP.
The D\O\ collaboration has not yet performed such a full mSUGRA 
scan. They investigated the configuration where the slepton mass is just above
$m_{\tilde\chi_2^0}$, along the edge of the right lobe, and obtained a chargino
mass limit of 145~GeV in that configuration~\cite{DZtril}. 
The gap between the two lobes 
corresponds to the case where two-body decays are just open, so that one of
the leptons arising from the $\tilde\chi_2^0\to\tilde\ell\ell$ decay is too
soft to be detected. This is the configuration where the search for same-sign
dileptons comes to rescue. A search for same sign dimuons was performed by 
D\O, which allows the gap to be reduced, as shown in Fig.~\ref{tril}-right. 

\subsection{Non-standard SUSY}

\subsubsection{Gauge mediated SUSY breaking}
In models where SUSY breaking is mediated by gauge interactions, the scale of
SUSY breaking is lower than in supergravity, and the gravitino becomes the LSP.
The phenomenology depends on the nature of the NLSP, the next to lightest SUSY
particle, and on its lifetime. In most viable models, the NLSP is a stau or a
neutralino. A stau NLSP was excluded at LEP with masses below 87 to 97~GeV,
depending on its lifetime~\cite{lepgmsb}. These remain the tighter limits to
date.

\begin{figure}
\begin{tabular}{cc}
\includegraphics[width=8.5cm]{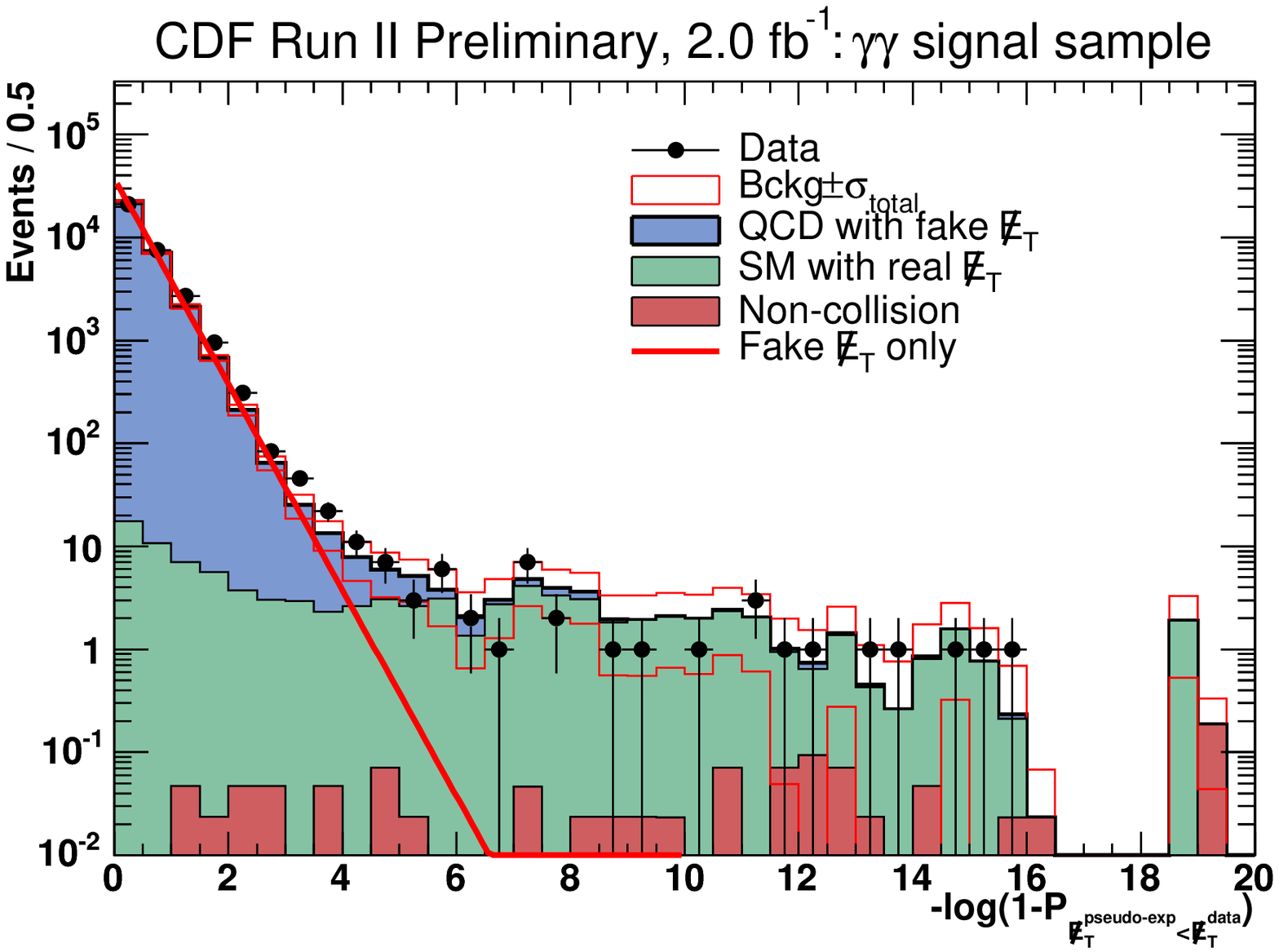} &
\includegraphics[width=8.5cm]{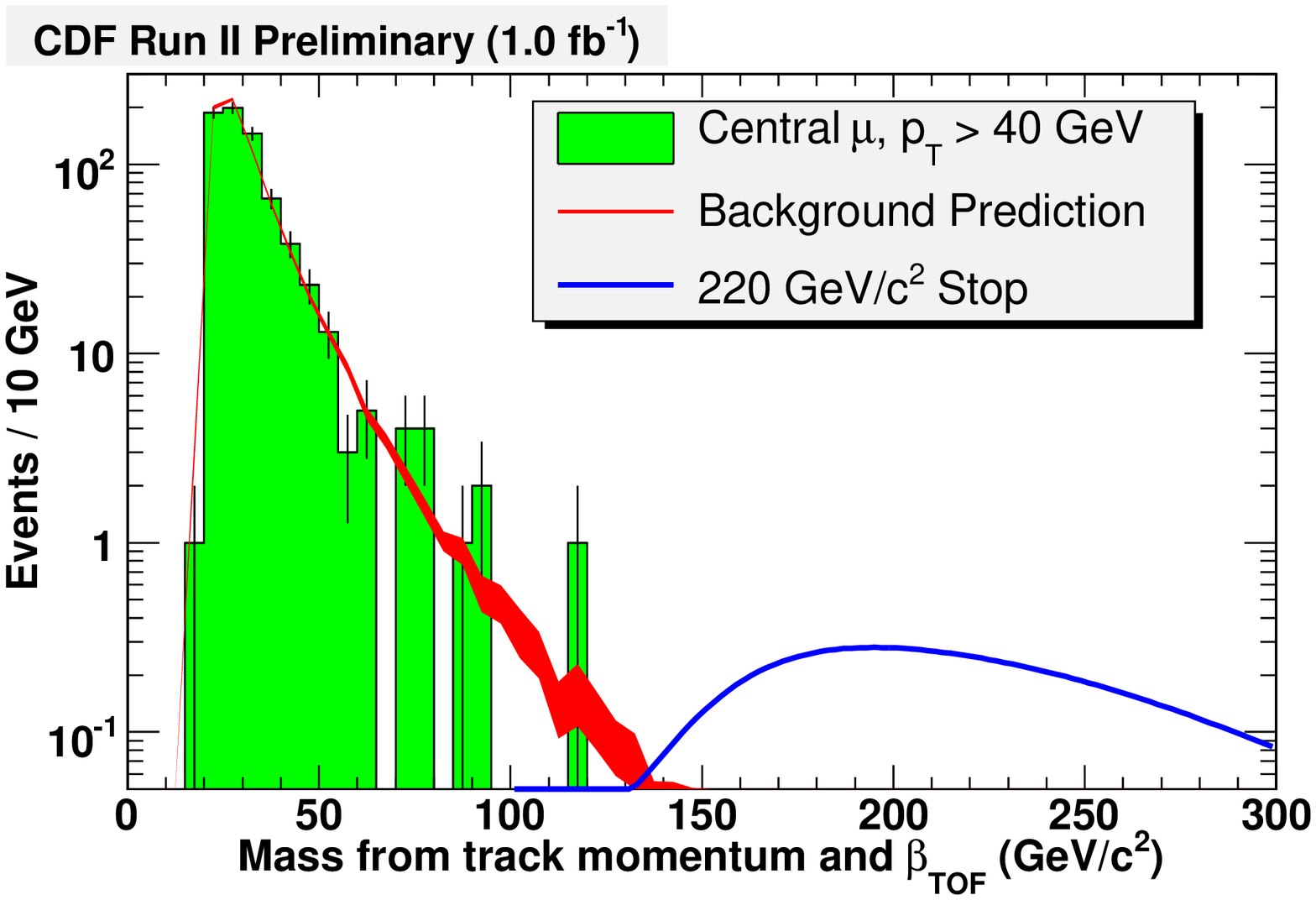}
\end{tabular}
\caption{\label{metsigcmsp}
Left: Distribution of the \met\ significance in the CDF search for diphotons
and \met~\cite{CDFgaga}.
Right: Distribution of the mass of charged particles, calculated by CDF from 
the momentum and time of flight~\cite{CDFchamp}.}
\end{figure}

In the mass range relevant for searches at LEP and at the Tevatron, a 
neutralino NLSP decays as $\tilde\chi_1^0\to\gamma\tilde{G}$, where the
very light gravitino $\tilde{G}$ escapes detection. The final state arising
from the pair production of SUSY particles therefore involves two photons and
missing energy carried away by the gravitinos. Inclusive searches for this
topology were performed at the Tevatron under both assumptions of prompt and
delayed NLSP decays. The distribution of the \met\ significance in the CDF
search with prompt decays is shown in 
Fig.~\ref{metsigcmsp}-left~\cite{CDFgaga}, 
where 
the resolutions on the energies of the objects involved in the calculation of 
the \met\ are folded into the significance. A clear separation is observed 
between processes with fake and with real \met. Based on an analysis of 
1.1~\invfb, the D\O\ collaboration interpreted their results within the 
framework of a minimal model, the ``Snowmass slope''~\cite{slope}, where the
dominant process is associated $\tilde\chi^\pm\tilde\chi_2^0$ production. A
chargino mass lower limit was set at 229~GeV~\cite{DZgmsb}.   

\subsubsection{Charged massive stable particles}
Long-lived charginos are expected in some models with anomaly-mediated SUSY 
breaking. The reason is that, in such models, the LSP is wino-like, so that the
$\tilde\chi^\pm - \tilde\chi_1^0$ mass difference is very small. Such 
long-lived charginos would appear in a Tevatron detector like slowly moving 
muons with large momenta. The D\O\ collaboration made use of the timing
information of their muon detectors to set a chargino mass lower limit of 
185~GeV, based on 1.1~\invfb~\cite{DZcmsp}. The CDF collaboration made use of
their time-of-flight detectors in 1~\invfb, 
as shown in Fig.~\ref{metsigcmsp}-right, and
excluded masses below 250~GeV for a long-lived stop~\cite{CDFchamp}, as would 
be the case in the MSSM for a small stop--LSP mass difference.

\section{EXTRA-DIMENSIONS}
Models with additional space dimensions come in many flavors: large, 
TeV$^{-1}$, universal, warped... The results reported in this review pertain
to the model of large extra-dimensions (LED) within which only gravity 
propagates in the bulk, and the  Kaluza-Klein (KK) excitations of the graviton
cannot be resolved. 

\begin{figure}
\begin{tabular}{cc}
\includegraphics[width=8.5cm]{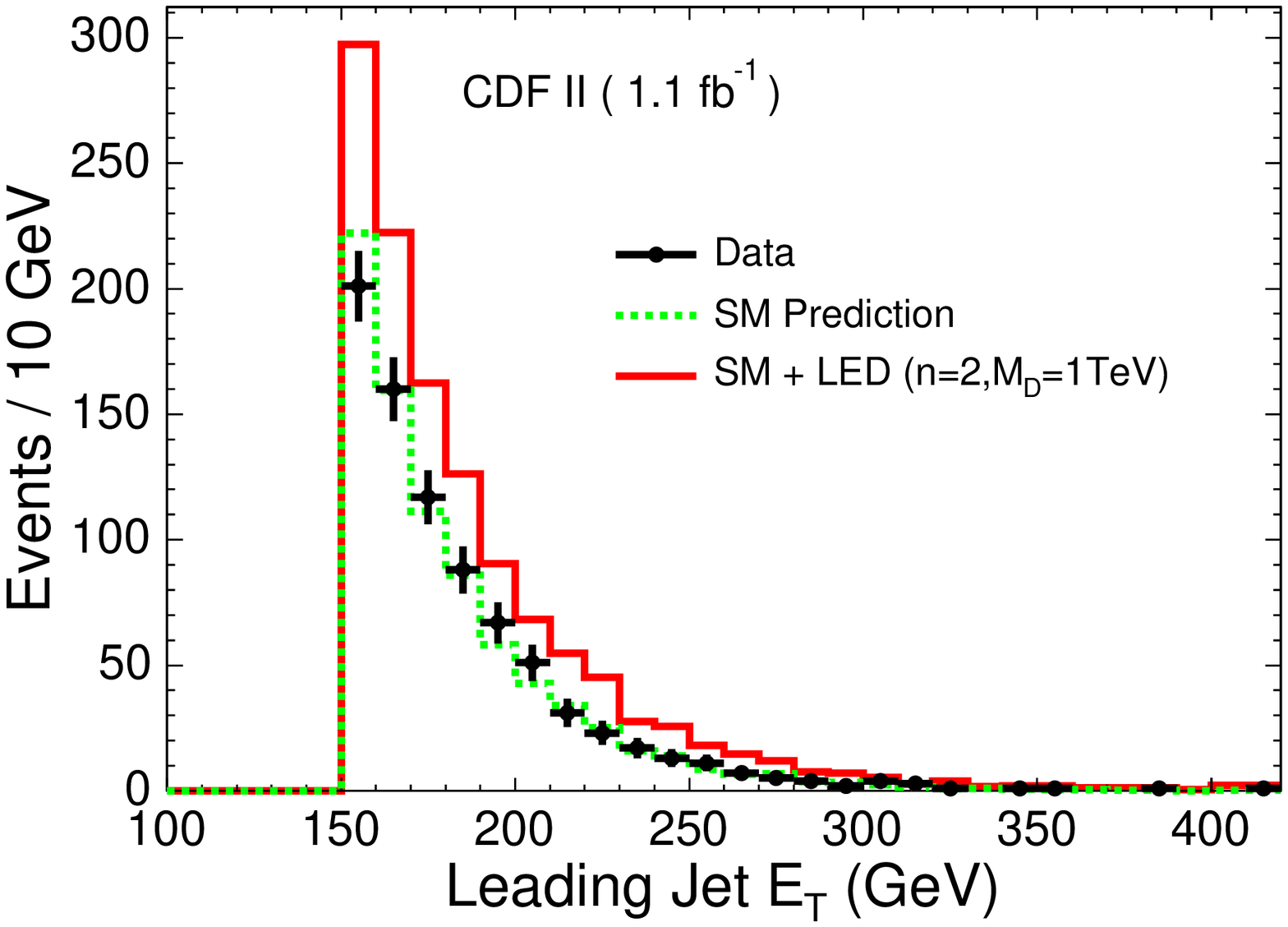} &
\includegraphics[width=8.5cm]{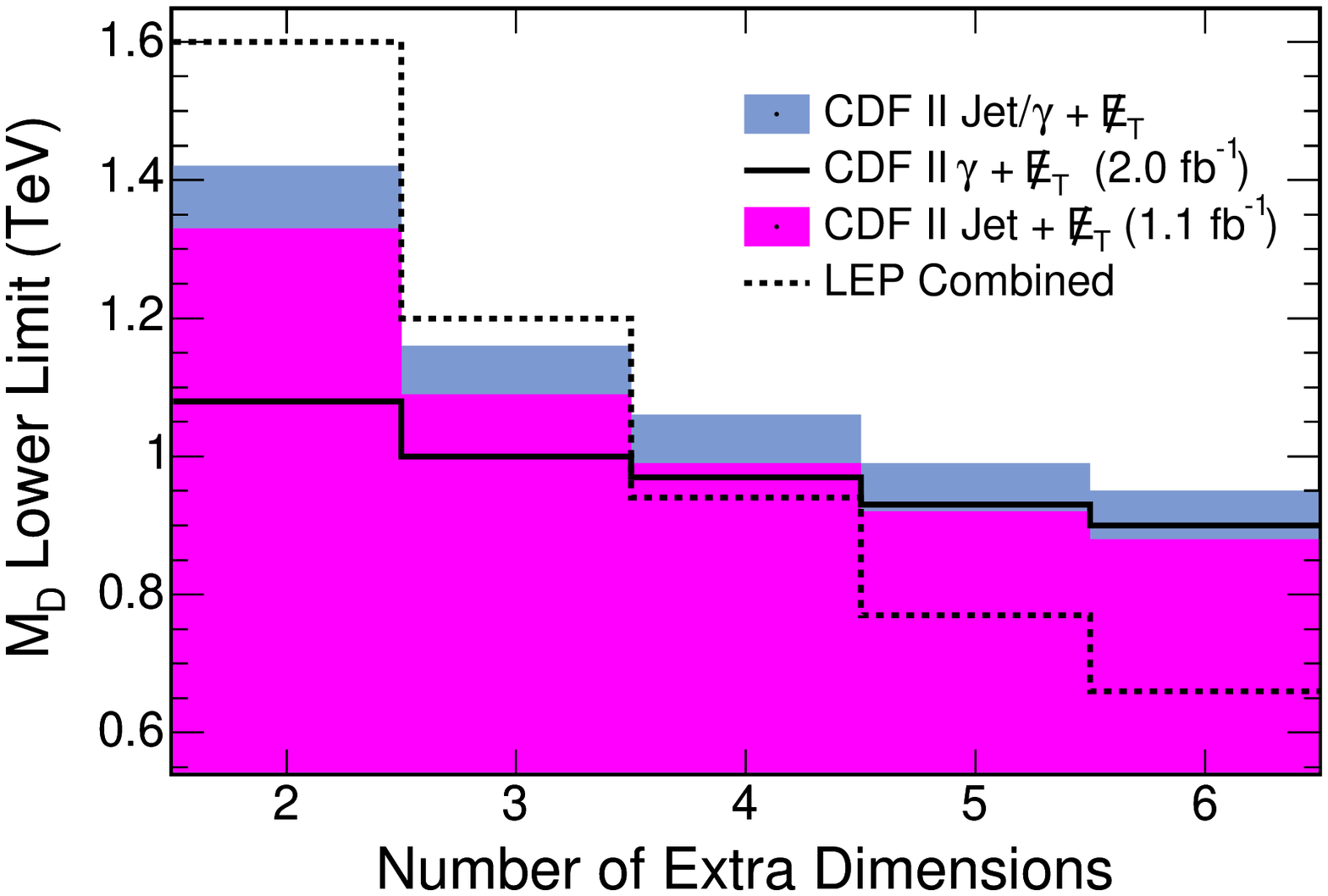}
\end{tabular}
\caption{\label{LED}
Left: Distribution of the jet $p_T$ in the CDF search for 
monojets~\cite{CDFLED}. 
Right: Limits on the fundamental Planck scale $M_D$ as a function of the number
of extra dimensions~\cite{CDFLED}.}
\end{figure}

Such KK gravitons can be produced together with a quark, a gluon, or a photon 
in $p\bar{p}$ collisions, and escape into the bulk. The corresponding final
states are a monojet or a mono-photon, with \met. The analysis of these 
processes provides direct information on the fundamental Planck scale $M_D$.
In the simplest model, $M_D$ is related to the Planck scale by 
$M_{Pl}^2=8\pi M_D^{n_D+2}R^{n_D}$, where $R$ is the radius of the $n_D$
additional compact dimensions.
A search for such monojets was performed in 1.1~\invfb\ by CDF~\cite{CDFLED}. 
As can be seen in Fig.~\ref{LED}-left, the spectrum
of jet $p_T$ for events with $\met>120$~GeV does not show any excess at large
$p_T$, above the SM expectation dominated by $(Z\to\nu\nu)$+jet, 
that could be attributed to the effects of LEDs. In addition to the SM 
$(Z\to\nu\nu)+\gamma$ background, the search for single photons suffers from 
a large background from cosmics and from the beam halo. The CDF 
analysis based on 2~\invfb~\cite{CDFLED} makes
use of the timing information of the calorimeter, while the D\O\ collaboration 
takes advantage of the fine segmentation of their electromagnetic calorimeter 
and of their preshower detector to verify that the photon candidates originate
from the interaction vertex~\cite{DZgamma}. The CDF collaboration combined
the results of their seraches for monojets and for single photons to extract
limits on $M_D$ as a function of the number of extra dimensions 
(Fig.~\ref{LED}-right). For $n_D=4$, $M_D>1.04$~TeV.

\begin{figure}
\begin{tabular}{cc}
\includegraphics[width=8.5cm]{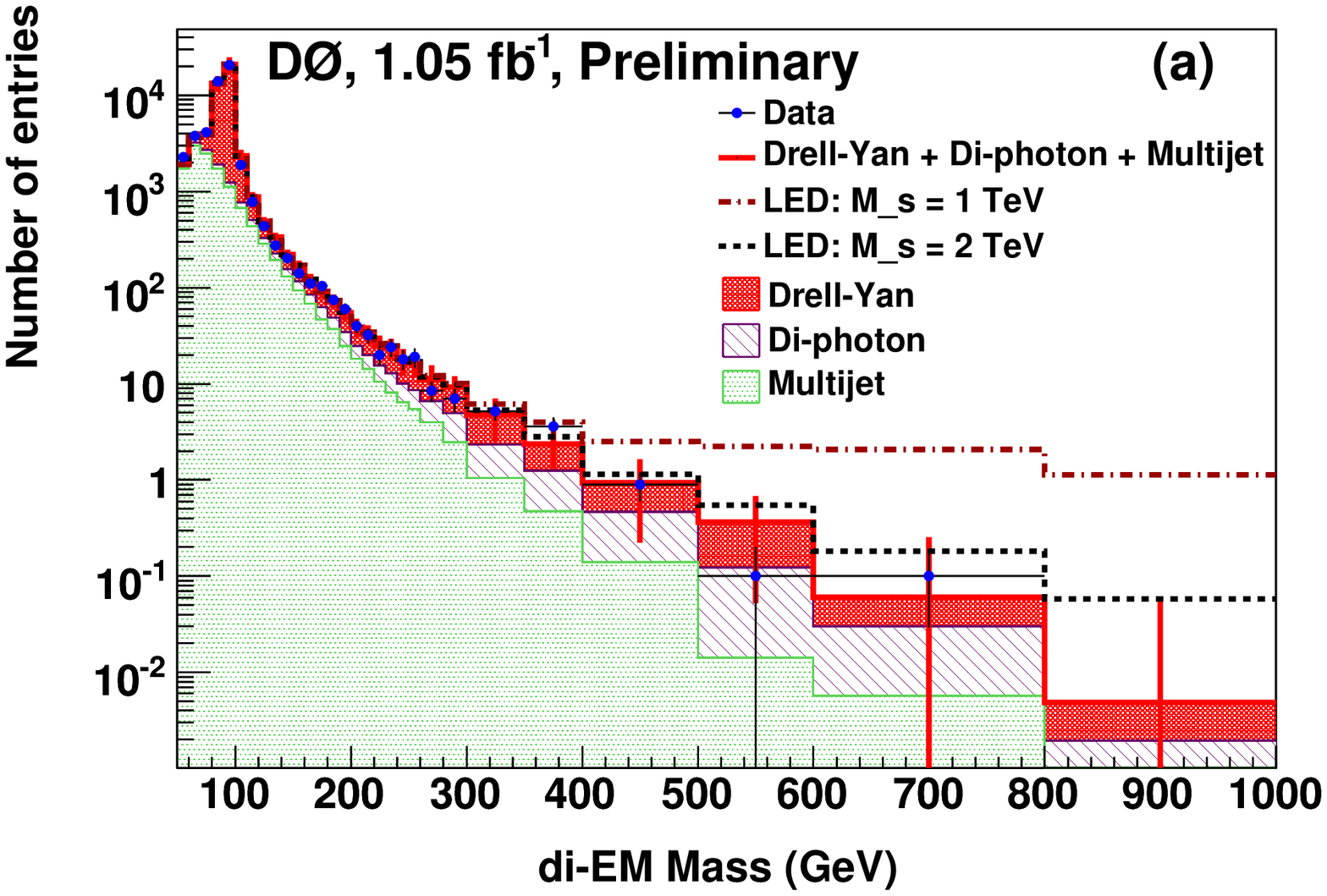} &
\includegraphics[width=8.5cm]{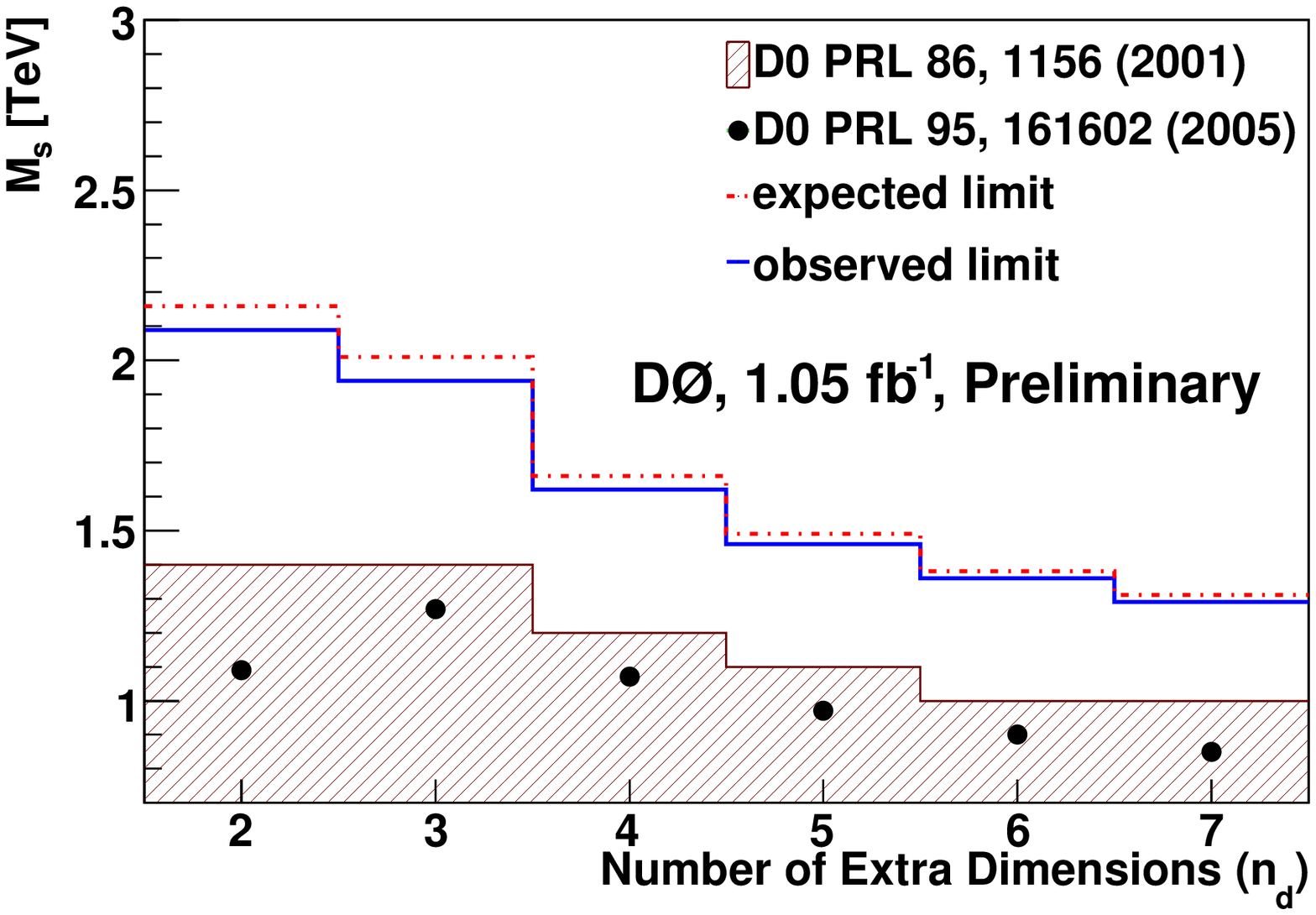}
\end{tabular}
\caption{\label{diEM}
Left: Distribution of the diEM mass in the D\O\ search for LEDs~\cite{DZLED}. 
Right: Limits on the effective Planck scale $M_s$ as a function of the number
of extra dimensions~\cite{DZLED}.}
\end{figure}

Virtual KK gravitons can also be exchanged in the production of fermion or
boson pairs, thus modifying the SM cross sections for these processes. Here, 
the sensitivity is expressed in terms of a cutoff, the effective Planck scale 
$M_s$, which is expected to be close to $M_D$. 
The D\O\ collaboration combined in a single analysis the search in 1.05~\invfb\
for LEDs in electron and photon pair production. The combined dielectron and 
diphoton (diEM) mass spectrum showed in Fig.~\ref{diEM}-left does not exhibit 
any excess at large diEM mass above the SM contribution, 
from which limits on $M_s$ are derived as a function of $n_D$ as shown in 
Fig.~\ref{diEM}-right~\cite{DZLED}. For $n_D=4$, $M_s>1.62$~TeV.

A search for LEDs in high $p_T$ dijets was performed by D\O~\cite{DZdijet} in 
0.7~\invfb. Given the uncertainties on the jet energy scale, the most sensitive
variable is the angular distribution in the center-of-mass of the hard process,
parameterized in terms of 
$\chi=$exp$(\vert y_1-y_2\vert)\sim (1+\cos\theta^\ast)/(1-\cos\theta^\ast)$,
where $y_1$ and $y_2$ are the rapidities of the two jets. The $\chi$
distribution would be uniform for Rutherford scattering, and the deviations 
are accurately predicted by perturbative QCD. Additional deviations are 
predicted in LED models, which would build up at large dijet masses and small 
values of $\chi$. The absence of such deviations allows a lower limit of 
1.49~TeV to be set on $M_s$ for $n_D=4$. 

\section{OTHER MODEL-INSPIRED SEARCHES}

\subsection{Extra gauge bosons}
\begin{figure}
\begin{tabular}{cc}
\includegraphics[width=8.5cm]{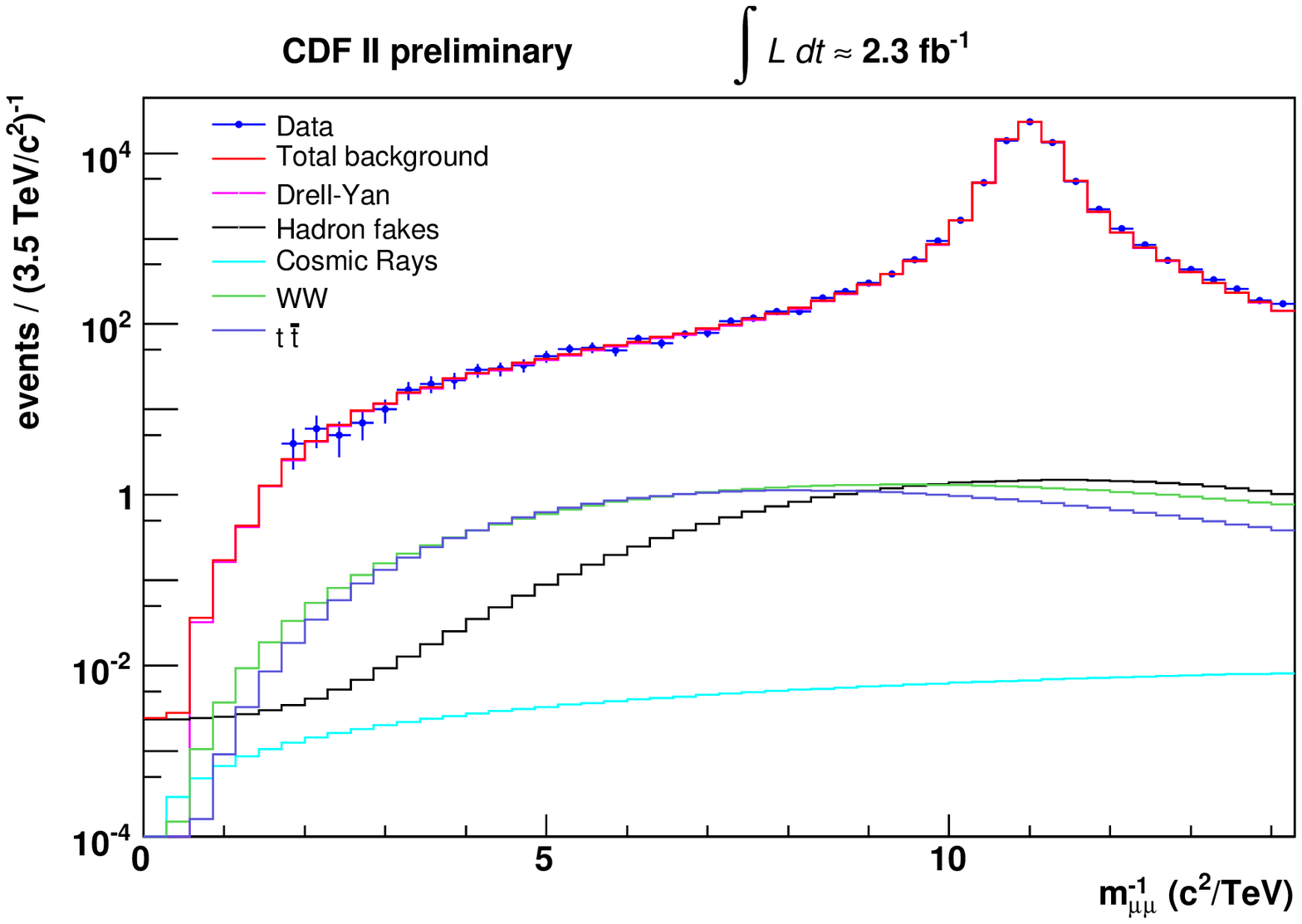} &
\includegraphics[width=8.5cm]{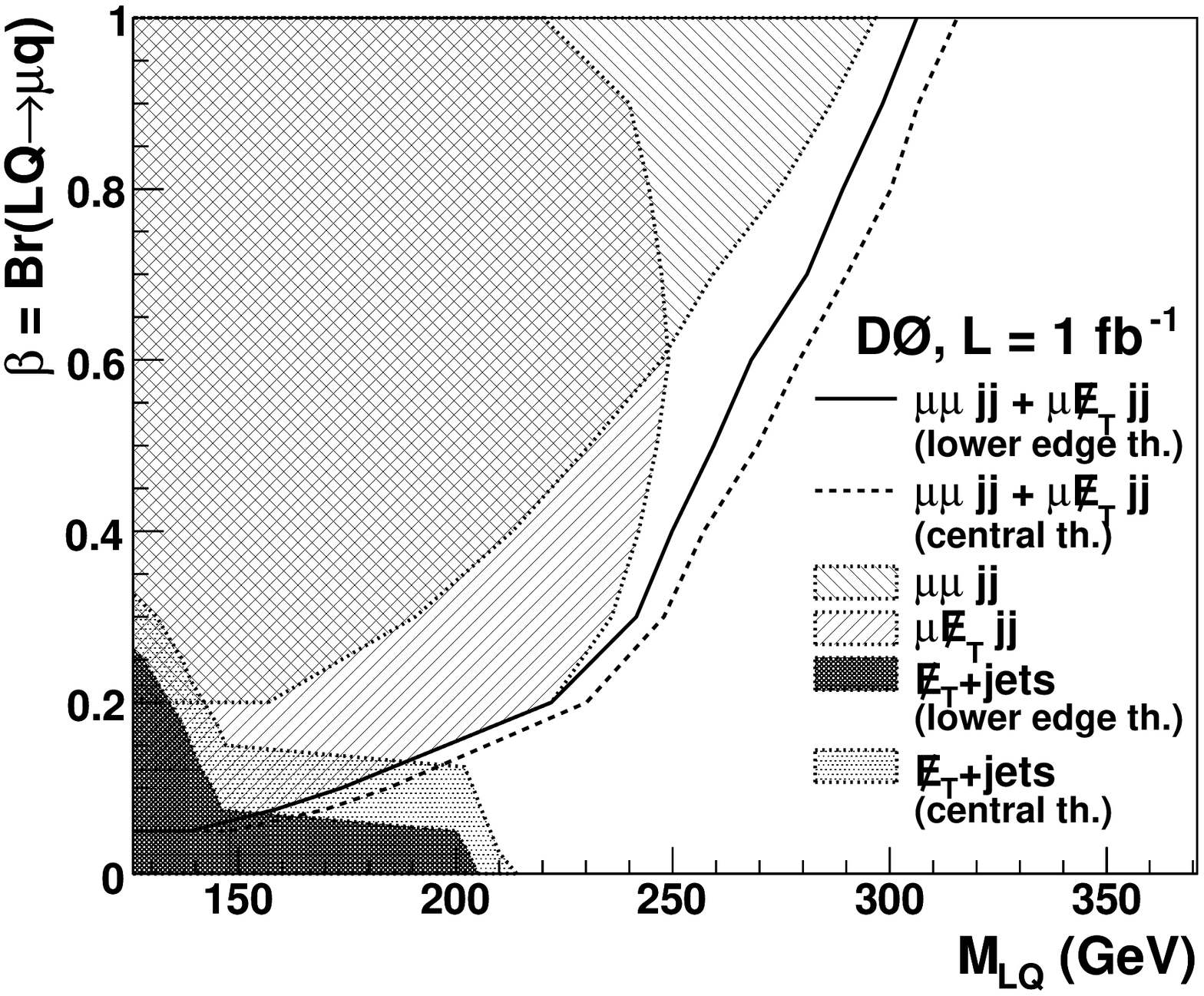}
\end{tabular}
\caption{\label{dimuLQ}
Left: Distribution of the dimuon inverse mass at CDF~\cite{CDFdimu}.
Right: Exclusion domain in the $(m_{LQ},\beta)$ plane from the D\O\ search
for second generation leptoquarks in the $\mu\mu qq$ and $\mu\nu qq$ 
topologies~\cite{DZsecond}; the low $\beta$ region is excluded by a search
in the acoplanar jet topology~\cite{DZLQ}.}
\end{figure}

Neutral bosons associated with extra U(1) gauge groups appear in many 
extensions of the standard model. Recent searches were performed by CDF for
such extra $Z'$ bosons in the dielectron~\cite{CDFdiel} and 
dimuon~\cite{CDFdimu} channels, based on 2.5 and 2.3~\invfb, respectively. 
They would show up as a peak in the dilepton 
mass distribution. The dimuon mass spectrum is shown in Fig.~\ref{dimuLQ}-left,
presented in terms of $1/m_{\mu\mu}$, a variable better fit to the trackers
resolution which is Gaussian in inverse momentum. In the
absence of any excess over the SM expectation, a lower limit of 1.03~TeV
was set on the mass of a sequential $Z'$, i.e., with couplings identical to 
those of the standard $Z$ boson. Somewhat lower limits were obtained
in various canonical E(6) models. In the dielectron channel, an excess is
seen around 240~GeV, the significance of which is however only 2.5$\sigma$,
once the trial factor is taken into account. 

Extra $W'$ bosons are also expected in extensions of the SM such as the
left-right symmetric model based on SU(2)$_L\times$SU(2)$_R\times$U(1). 
A D\O\ search for $W'\to e\nu$ in 1~\invfb\ excluded a sequential $W'$ 
with mass below 1.00~TeV~\cite{DZWprime}. 

\subsection{Leptoquarks}
The apparent symmetry between the structures of the lepton and quark
representations in the SM leads to the conjecture that bosons carrying
lepton and quark quantum numbers could mediate transitions between quarks and
leptons. Indeed, such leptoquarks appear naturally in, e.g., grand unified 
theories. In order to avoid flavor changing neutral currents, it is commonly
assumed that a given leptoquark couples to a single generation. Leptoquarks
decay into a quark and a charged or neutral lepton, with branching fractions
$\beta$ and $(1-\beta)$, respectively. In the following, only scalar
leptoquarks are considered.

At the Tevatron, leptoquarks can be pair produced by the strong interaction,
$p\bar{p}\to LQ \overline{LQ}$. Depending on the way they decay, the final 
state is $\ell^+\ell^- q\bar{q}$, $\ell\nu q\bar{q}$, or $\nu\bar\nu q\bar{q}$,
with respective branching fractions $\beta^2$, $2\beta(1-\beta)$, or 
$(1-\beta)^2$. Searches were recently performed by D\O\ for leptoquarks of all
three generations. The search for first generation leptoquarks was performed
in the $e^+e^- q\bar{q}$ channel, from which a mass lower limit of 299~GeV was
derived for $\beta=1$~\cite{DZfirst}. The search for third generation 
leptoquarks was performed in the $\tau^+\tau^- b\bar{b}$ channel, with one 
$\tau$ decaying as $\tau\to\mu\nu\nu$, and the other as 
$\tau\to$~hadron(s)+$\nu$. A mass lower limit of 210~GeV was obtained for 
$\beta=1$~\cite{DZthird}. The search for second generation leptoquarks was
performed in both the $\mu^+\mu^- q\bar{q}$ and $\mu\nu q\bar{q}$ channels,
thus allowing an exclusion domain to be determined in the $(m_{LQ},\beta)$
plane, as shown in Fig.~\ref{dimuLQ}-right~\cite{DZsecond}. For $\beta=1$,
a mass lower limit of 316~GeV was obtained. All these searches were performed
in a data sample of 1~\invfb. A search for acoplanar jets, relevant for 
$\beta=0$, was performed in 2.5~\invfb~\cite{DZLQ}. It is similar to the 
search for squarks reported earlier in this review, and allowed first and 
second-generation leptoquarks to be excluded with masses below 214~GeV for 
$\beta=0$. The contribution of this search at low $\beta$ is shown in 
Fig.~\ref{dimuLQ}-right. A tighter limit of 229~GeV was obtained previously 
for third generation leptoquarks decaying into $\nu\bar\nu b\bar{b}$, in which
case heavy flavor tagging increases the sensitivity~\cite{DZLQ3}.

\begin{figure}
\begin{tabular}{cc}
\includegraphics[width=8.5cm]{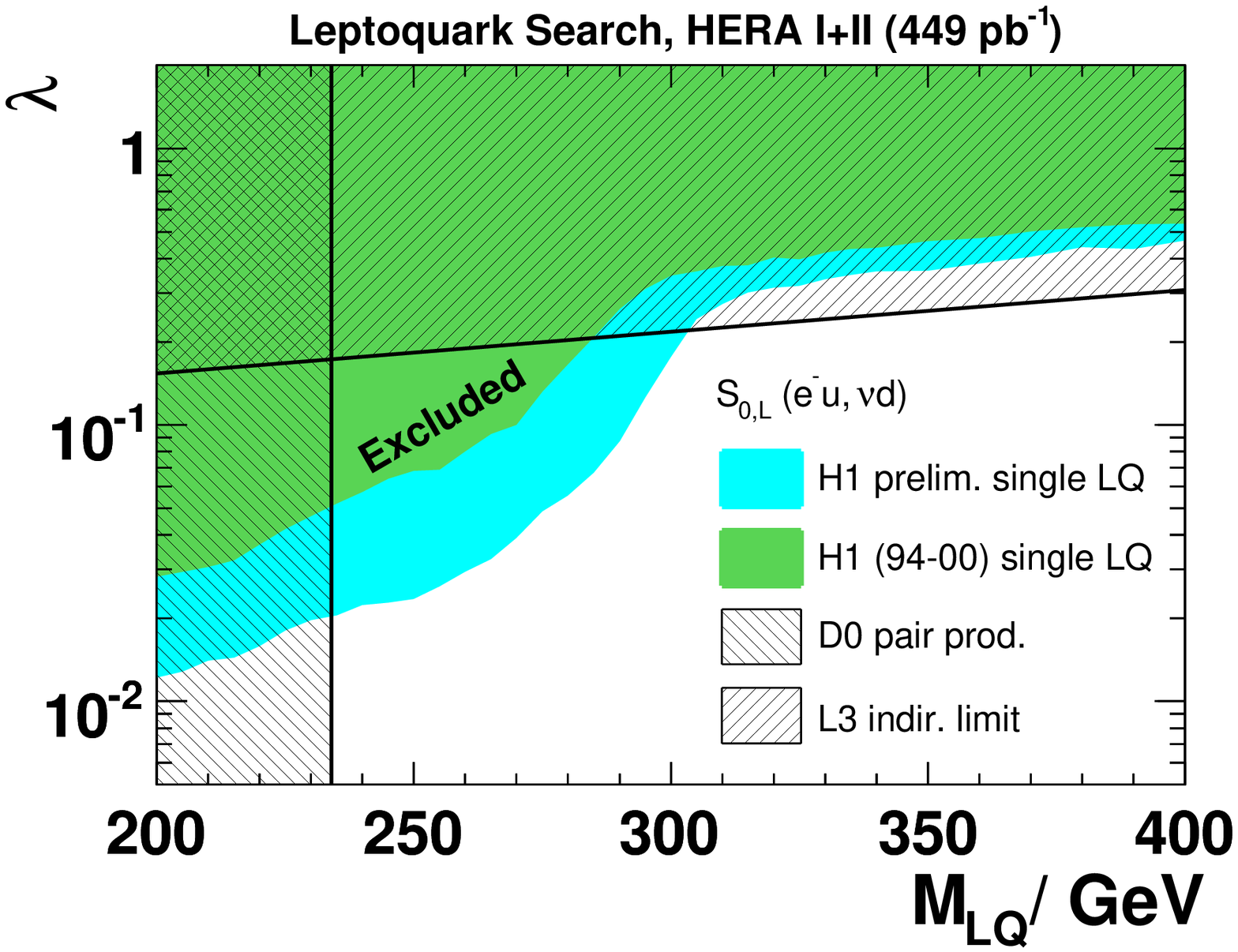} &
\includegraphics[width=8.5cm]{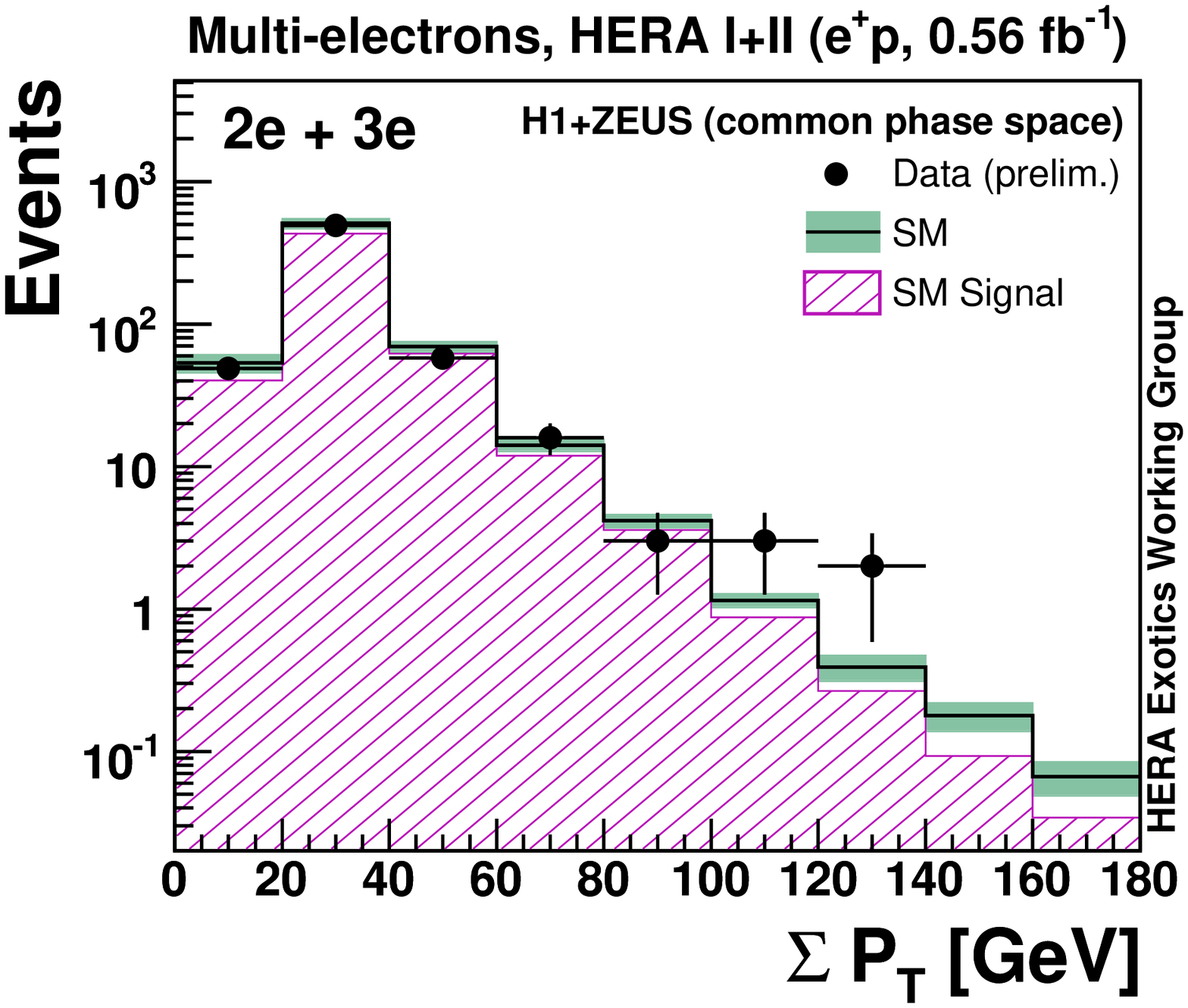}
\end{tabular}
\caption{\label{HERA}
Left: Domain excluded in the $(m_{LQ},\lambda)$ plane by the H1 search for
first generation leptoquarks with $\beta=0.5$~\cite{LQH1}. 
Right: Distribution of the scalar sum of the transverse momenta of the 
electrons in the two and three-electron final states in $e^+p$ collisions 
at HERA (combined H1 and ZEUS result)~\cite{H1Zmulti}.}
\end{figure}

First generation leptoquarks can also be resonantly produced in $ep$ collisions
at HERA, $eq\to LQ$. The production cross section is however dependent on the
a priori unknown $eqLQ$ coupling $\lambda$, so that the resulting mass limit
depends on the value of $\lambda$. Searches were performed by 
H1~\cite{LQH1} in the $eq$ and
$\nu q$ final states in 0.45~\invfb. The excluded domain obtained for $\beta=1$
is almost entirely covered by the search for first generation leptoquarks 
reported above, and by the indirect limits from LEP. For $\beta=0.5$, however,
a substantial domain is excluded only by the H1 analysis, as shown in 
Fig.~\ref{HERA}-left.
 
\section{SEARCHES FOR THE UNEXPECTED}
Up to here, the searches reported were all motivated by some theoretical model.
It is also possible to look for deviations from the standard model, without 
any specific a priori motivation.

\subsection{Anomalies at HERA}
In the HERA~I data taking period, the H1 collaboration observed an excess of
events with an isolated lepton, missing transverse energy, and a high $p_T$
hadronic jet, in $e^+p$ collisions only. 
The SM process leading to this topology is the photoproduction 
of a $W$ boson, $\gamma q\to q' W$, with $W\to\ell\nu$. The full HERA I+II
statistics was reanalysed by both H1 and ZEUS, with similar selection 
criteria~\cite{H1Zanomaly}. For $p_T>25$~GeV, 
an excess is still observed by H1 in their $e^+p$ data, 17 events for 
$7.1\pm 0.9$ expected, while ZEUS selects six events for $7.5\pm 1.1$ expected.
The significance of the excess in the combination of the two experiments, 
23 events vs. $14.6\pm 1.9$ expected, is only at the level of $1.8\sigma$.

An excess was also observed by H1 in the topology with two or three high $p_T$
leptons. The SM process leading to this final state is the two-photon 
interaction $\gamma\gamma\to\ell^+\ell^-$, with possibly in addition the beam 
electron deflected at large angle. In their full HERA~I+II $e^+p$ dataset, five
events were observed by H1 with two or three electrons or muons and with a sum 
of lepton transverse momenta larger than 100~GeV, while only $0.96\pm 0.12$
were expected~\cite{H1multi}. No such excess was seen in the $e^-p$ data. The
H1 and ZEUS collaborations agreed on a common selection to combine their
results in the multilectron channel~\cite{H1Zmulti}. 
For $\Sigma p_T$(leptons)$>100$~GeV, they observe five events
in $e^+p$ collisions, for $1.8\pm 0.2$ expected (Fig.~\ref{HERA}-right). 
In $e^-p$ collisions, one
event is selected, in agreement with the $1.2\pm 0.1$ expected. The full
combination, including final states with muons, is currently underway.

\subsection{Global, model-independent analysis}
In a model-independent approach all events are categorized in terms of their 
content
in high $p_T$ objects: electrons, muons, $\tau$ leptons, photons, jets, 
$b$ jets, and neutrinos, i.e. \met. A comparison is then performed between the
numbers of events in each class and the expectations from all SM processes.
Such an approach was pioneered by H1 at HERA~I, who repeated their analysis 
with the HERA~II dataset. Excellent agreement was observed, except for the 
known excess in the $\mu\nu$+jet final state in $e^+p$ collisions~\cite{H1MIS}.

The CDF collaboration considered 399 final states in a 2~\invfb\ data sample.
After adjusting a few normalization and correction factors, no significant
anomaly was uncovered. A total of 19650 kinematic distributions was next
analysed using the {\sc vista} package, 
of which a few hundred were found to be discrepant with respect to 
their SM expectation.
All these discrepancies could 
however be traced to an inadequate modeling of soft QCD effects in the 
simulation, and none was found to be suggestive of any new physics. 
They next proceeded to investigate, with the {\sc sleuth} algorithm, 
more specifically the high $p_T$ tails, where new 
physics is expected to be more likely to show up. They also performed a search 
for mass bumps. Taking into account the trial factor, no excess was found
beyond what is expected from statistics~\cite{CDFMIS}. A similar investigation
was performed by H1, reaching the same conclusion~\cite{H1MIS}.

Although the sensitivity of such a model-independent approach does not 
compete with that of dedicated searches, it is important to make sure that 
no stone was left unturned, in case the right new physics model has not yet 
been considered...

\section{FINAL REMARKS}

One year after the end of the HERA~II run, the H1 and ZEUS collaborations are
delivering their final results, including combinations corresponding to an
integrated luminosity of almost 1~\invfb. A couple of intriguing anomalies
in the H1 $e^+p$ data will most likely remain with us in the foreseeable
future.

Broad searches for phenomena beyond the standard model are being pursued by the
CDF and D\O\ collaborations at the Tevatron, with steadily increasing 
integrated luminosities. Currently, results are reported by the individual 
experiments with data samples corresponding to up to 2.5~\invfb. 
Unfortunately, no sign of new physics has been uncovered until now. With
4~\invfb\ on tape per experiment, it is time to update the existing searches,
and to reach a sensitivity corresponding to 8~\invfb\ by combining the CDF 
and D\O\ results. And with the LHC imminent start-up, time is
running short for the Tevatron collaborations... 



\end{document}